\documentclass[twocolumn]{aastex631}

\usepackage{CJK}
\usepackage[whole]{bxcjkjatype} 
\usepackage{CJKutf8}
\usepackage{threeparttable}

\begin{document}

\title{SN\,2021ukt: A Transitional Supernova with a Short Plateau and Persistent Interaction}

\author[0009-0009-7665-6827]{Neil R. Pichay}
\affiliation{Department of Astronomy, University of California, Berkeley, CA 94720-3411, USA}
\email{14neil@berkeley.edu}

\author[0000-0002-4951-8762]{Sergiy S. Vasylyev}
\affiliation{Department of Astronomy, University of California, Berkeley, CA 94720-3411, USA}
\affiliation{Department of Astronomy \& Astrophysics, University of California, San Diego, 9500 Gilman Drive, MC 0424, La Jolla, CA 92093-0424, USA}

\author[0009-0008-2961-4328]{Audrey M. Liddle}
\affiliation{Department of Astronomy, University of California, Berkeley, CA 94720-3411, USA}

\author[0000-0003-3460-0103]{Alexei V. Filippenko}
\affiliation{Department of Astronomy, University of California, Berkeley, CA 94720-3411, USA}

\author[0000-0002-2636-6508]{WeiKang Zheng}
\affiliation{Department of Astronomy, University of California, Berkeley, CA 94720-3411, USA}

\author[0000-0001-5955-2502]{Thomas G. Brink}
\affiliation{Department of Astronomy, University of California, Berkeley, CA 94720-3411, USA}

\author[0000-0002-6535-8500]{Yi Yang
\begin{CJK}{UTF8}{gbsn}
(杨轶)
\end{CJK}}
\affiliation{Department of Physics, Tsinghua University, Qinghua Yuan, Beijing 100084, China}
\email{yi\_yang@mail.tsinghua.edu.cn}

\author{Matthew Graham}
\affiliation{Division of Physics, Mathematics, and Astronomy, California Institute of Technology, Pasadena, CA 91125, USA}

\author[0000-0003-2686-9241]{Daniel Stern}
\affiliation{Jet Propulsion Laboratory, California Institute of Technology, 4800 Oak Grove Drive, Pasadena, CA 91109, USA}

\author{Daichi Hiramatsu}
\affiliation{Department of Astronomy, University of Florida,
211 Bryant Space Science Center, Gainesville, FL 32611-2055 USA}
\affiliation{Center for Astrophysics | Harvard \& Smithsonian, 60 Garden Street, Cambridge, MA 02138-1516, USA}
\affiliation{The NSF AI Institute for Artificial Intelligence and Fundamental Interactions, USA}

\author[0000-0003-2375-2064]{Claudia P. Guti\'{e}rrez}
\affiliation{Institute of Space Sciences (ICE, CSIC), Campus UAB, Carrer de
Can Magrans, s/n, E-08193 Barcelona, Spain}
\affiliation{Institut d'Estudis Espacials de Catalunya (IEEC), 08860 Castelldefels (Barcelona), Spain}

\author{K. Azalee Bostroem}
\affiliation{Steward Observatory, University of Arizona, 933 North Cherry Avenue, Tucson, AZ 85721-0065, USA}
\altaffiliation{LSST-DA Catalyst Fellow}


\author{Estefania Padilla Gonzalez}
\affiliation{Space Telescope Science Institute, 3700 San Martin Drive, Baltimore, MD 21218-2410, USA}

\author[0000-0003-4253-656X]{D. Andrew Howell}
\affiliation{Las Cumbres Observatory, 6740 Cortona Drive, Suite 102, Goleta, CA 93117-5575, USA}
\affiliation{Department of Physics, University of California, Santa Barbara, CA 93106-9530, USA}

\author{Curtis McCully}
\affiliation{Las Cumbres Observatory, 6740 Cortona Drive, Suite 102, Goleta, CA 93117-5575, USA}

\author{Megan Newsome}
\affiliation{Department of Astronomy, University of Texas at Austin, 2515 Speedway, Stop C1400, Austin, TX 78712, USA}

\author{Craig Pellegrino}
\affiliation{Department of Astronomy, University of Virginia, Charlottesville, VA 22903, USA}
\affiliation{Goddard Space Flight Center, 8800 Greenbelt Road, Greenbelt, MD 20771, USA}

\author{Giacomo Terreran}
\affiliation{Adler Planetarium, 1300 S. DuSable Lake Shore Dr., Chicago, IL 60605, USA}

\author{Ivan Altunin}
\affiliation{Department of Physics, University of Nevada, Reno, NV 89557, USA}

\author[0009-0004-7268-7283]{Raphael Baer-Way}
\affiliation{Department of Astronomy, University of Virginia, Charlottesville, VA 22904, USA}

\author{Vidhi Chandler}
\affiliation{Department of Astronomy, University of California, Berkeley, CA 94720-3411, USA}

\author{Asia A. deGraw}
\affiliation{Department of Astronomy, University of California, Berkeley, CA 94720-3411, USA}

\author{Connor F. Jennings}
\affiliation{Department of Astronomy, University of California, Berkeley, CA 94720-3411, USA}

\author{Michael B. May}
\affiliation{Department of Astronomy, University of California, Berkeley, CA 94720-3411, USA}
\affiliation{Department of Nuclear Engineering, University of Tennessee, Knoxville, TN 37996, USA}

\begin{abstract}
We present spectroscopic and photometric observations of supernova (SN)\,2021ukt, a peculiar short-plateau object that was originally identified as a Type IIn SN and later underwent an unprecedented transition to a Type Ib (possibly Type IIb) SN. The early-time light curves of SN\,2021ukt exhibit a $\sim 25$\,day plateau. Such a short phase of hydrogen recombination suggests a rather thin H-rich outer envelope of the progenitor star. The relatively narrow Balmer emission lines in  spectra of SN\,2021ukt during the first week indicate the interaction between the expanding ejecta and the immediate circumstellar material (CSM). This H$\alpha$ line is observed throughout its helium-rich ejecta-dominated phase and nebular phase, suggesting persistent interaction with a radially extended CSM profile. We explore the synthetic light-curve model among grids of parameters generated by \texttt{MESA+STELLA}. We also compare the spectrophotometric evolution of SN\,2021ukt with several well-sampled supernovae that exhibit a short plateau and persistent ejecta-CSM interaction. An estimate of the progenitor mass of SN\,2021ukt is made based on the flux ratio between [\ion{Ca}{2}] $\lambda \lambda$7291, 7324 and [\ion{O}{1}] $\lambda \lambda$6300, 6364 during its nebular phase. Our analysis suggests that the progenitor star of SN\,2021ukt has a zero-age main-sequence (ZAMS) mass of $M_{\text{ZAMS}}\approx 12\,\text{M}_{\odot}$, a mass of radioactive $^{56}$Ni synthesized in the SN ejecta of $M_{\text{Ni}}\lesssim0.04\,\text{M}_{\odot}$, and a mass of the H-rich envelope of $M_{\rm H,env}\lesssim0.5\,\text{M}_{\odot}$. This study adds to the growing sample of transitional supernovae, reinforcing evidence for a continuum of underrepresented progenitors whose evolutionary pathways lie between those of standard SN models.

\end{abstract}


\section{Introduction}\label{sec:intro}
When the core of a massive star ($M >8\,\text{M}_{\odot}$) consists of iron and is no longer able to undergo nuclear fusion, it collapses under its own gravity to form a protoneutron star, and the stellar envelope falls inward. 
The protoneutron star rebounds,  sending a shock wave outward that triggers an explosion of the stellar envelope, propelled mostly by neutrino interactions.  This process leads to a core-collapse (CC) supernova (SN). Among all known CC supernovae (SNe), about 63\% are found to be hydrogen-rich Type II supernovae~\citep[hereafter SNe\,II; e.g.,][]{2009ARA&A..47...63S, 2011MNRAS.412.1522S}.
The observed diversity among H-rich CCSNe can be attributed to differences in the mass of the progenitor star and the mass of the hydrogen envelope at the moment of explosion; the latter can also be influenced by mass loss and binary interactions~\cite[][and references therein]{nomoto1995evolution, Eldridge_Xiao_Stanway_Rodrigues_Guo_2018, 10.1111/j.1365-2966.2011.18598.x}. 

Immediately after the initial rise in the light curve, SNe\,IIP exhibit a distinctive $\sim 80$--120 day plateau.  SNe\,IIL are observed with a characteristic linear decline in magnitude. SNe\,IIb display spectra with initially H-dominant to later He-dominant phases in the first month after the SN explosion \citep{1993ApJ...415L.103F}, indicating a thinner hydrogen envelope compared to other Type II subtypes. SNe\,IIn exhibit evidence of interaction between the ejecta and the circumstellar material (CSM) through relatively narrow Balmer emission lines \citep{1990MNRAS.244..269S}. Based on pre-explosion imaging and stellar-evolution models, the likely progenitor scenarios are red supergiants (RSGs) with masses in the range 10--17\,$\text{M}_{\odot}$ and extended H-rich envelopes for SNe\,IIP~\citep{1976ApJ...207..872C,1977ApJS...33..515F}; yellow supergiants (YSGs) with a binary companion for SNe\,IIb~\citep{1993ApJ...415L.103F, 1994ApJ...429..300W, maund2004massive}; and luminous blue variable stars undergoing enhanced mass loss for SNe\,IIn~\citep{humphreys1994luminous, gal2009massive, smith2011massive}. The progenitors of SNe\,IIL are not yet well determined, but are possibly RSGs or YSGs (see~\citealt{2017hsn..book..693V} for review). Among the wide diversity of H-rich CCSNe, detailed investigations are establishing links between various subtypes. For example, \cite{2014ApJ...786...67A} suggest that
a continuum of pre-explosion mass-loss history and evolution, primarily influenced by the amount of H retained by the progenitor star, may account for the transition from the no plateau (SNe\,IIL) to short and long plateau SN\,II subtypes. Historically, observations of short plateaus (lasting $\sim 20$--70 days after first light) remain scarce \citep{2021ApJ...913...55H}. 

\begin{table*}[ht]
\centering
\begin{threeparttable}
    \caption{Log of Spectroscopic Observations}
    \label{table:spec_log}
    \begin{tabular}{l|c|l|c|c}
        \hline
        \textbf{Obs. Date, Time}$^{\text{a}}$ &
        \textbf{Phase}$^{\text{b}}$ &
        \textbf{Facility/Instr.} &
        \textbf{Exp. (s)} &
        \textbf{Range (\AA)} \\
        \hline

        2021-08-01 13:49:59 & 3   & UH88/SNIFS   & 1500 & 3400--9100   \\
        2021-08-03 11:41:04 & 5   & LCO/FLOYDS-N & 1200 & 3500--10000  \\
        2021-08-06 12:17:32 & 8   & LCO/FLOYDS-N & 1200 & 3500--10000  \\
        2021-08-09 11:57:22 & 11  & LCO/FLOYDS-N & 1200 & 3500--10000  \\
        2021-08-15 13:30:07 & 17  & LCO/FLOYDS-N & 1200 & 3500--10000  \\
        2021-08-19 14:38:17 & 21  & LCO/FLOYDS-N & 1200 & 3500--10000  \\
        2021-08-27 15:10:45 & 29  & LCO/FLOYDS-S & 1200 & 3500--10000  \\
        2021-08-31 08:16:48 & 33  & Lick/Kast    & 1500 & 3622--10748  \\
        2021-08-31 10:25:41 & 33  & LCO/FLOYDS-N & 1200 & 3500--10000  \\
        2021-09-04 11:15:22 & 37  & Lick/Kast    &  752 & 3624--10750  \\
        2021-09-04 12:08:46 & 37  & LCO/FLOYDS-N & 1200 & 3500--10000  \\
        2021-09-08 10:55:29 & 41  & Keck/LRIS    &  450 & 3137--10277  \\
        2021-09-09 12:43:45 & 42  & LCO/FLOYDS-S & 2700 & 3500--10000  \\
        2021-09-11 06:56:10 & 44  & Lick/Kast    & 2400 & 3634--10756  \\
        2021-09-18 10:51:57 & 51  & LCO/FLOYDS-N & 3600 & 3500--10000  \\
        2021-09-28 10:04:48 & 61  & Lick/Kast    & 3600 & 3632--10610  \\
        2021-09-30 09:59:09 & 63  & LCO/FLOYDS-N & 3600 & 3500--10000  \\
        2021-10-09 10:46:54 & 72  & LCO/FLOYDS-S & 3600 & 3500--10000  \\
        2021-10-15 08:25:26 & 78  & Lick/Kast    & 3600 & 3622--10538  \\
        2022-01-25 05:36:13 & 180 & Keck/LRIS    & 3000 & 3140--10327  \\
        \hline
    \end{tabular}

    \begin{tablenotes}
        \centering
        \footnotesize
        \item[$^{\text{a}}$] Date in UTC.
        \item[$^{\text{b}}$] Days since the estimated time of first light at MJD 59424 $\pm$ 2.5.
    \end{tablenotes}

\end{threeparttable}
\end{table*}

Some progenitors can lose their outer H and/or He envelope and explode as H-poor or even H-free stripped-envelope CCSNe (SESNe); for reviews, see~\cite{1997ARA&A..35..309F} and~\cite{2017hsn..book..277P}. The subclasses of SESNe include H-poor SNe\,IIb, H-free (or highly H-poor) SNe\,Ib, and H-free + He-poor/free SNe\,Ic~\citep{1994ApJ...429..300W}. The progenitors of SESNe are likely low-mass stars in binary systems (\citealp{nomoto1995evolution, 10.1093/mnras/sty3399, 2019NatAs...3..717M}). The envelope stripping can be caused by both stellar winds from sufficiently massive stars~\citep{10.1111/j.1365-2966.2008.14134.x} and (more frequently) binary interactions~\citep{2014ARA&A..52..487S}. Some studies suggest that these envelopes are stripped in the late stages of stellar mass loss (see, e.g.,~\citealp{2014ApJ...781...42O, Strotjohann_2021}).

Observations of CCSNe have shown signatures of interaction with pre-existing CSM. The relatively narrow or intermediate-width H$\alpha$ emission (full width at half-maximum intensity (FWHM) $\approx$ 1000\,km\,s$^{-1}$) observed in SNe\,IIn (e.g., SN\,2010jl;~\citealt{2014ApJ...781...42O}) is found to be the result of interaction between the ejecta and dense, optically thick, H-rich CSM~\citep{2000ApJ...536..239L, 10.1093/mnras/stac1093}. Furthermore, late-time observations months after the explosion can reveal interaction with the ejecta and an extended CSM shell (see, e.g., \citealp{2000AJ....120.1499M, 2018MNRAS.477...74A}). One peculiar case of CSM interaction was with SN\,2014C, which initially behaved as a normal SN\,Ib. After $\sim 100$ days, the spectrum revealed relatively narrow H$\alpha$ emission. This was found to be partly due to delayed interaction with CSM formed by the material stripped from the star~\citep{2015ApJ...815..120M}.

Here we present optical spectroscopy and photometry of SN\,2021ukt. The early-time spectra of SN\,2021ukt display a close resemblance to features of SNe\,IIn, and later observations reveal features resembling those of SNe\,Ib with evidence of CSM interaction. Meanwhile, the photometric evolution of SN\,2021ukt suggests that it belongs to the growing population of short-plateau SNe. In Section~\ref{sec:obs}, we discuss the discovery of SN\,2021ukt, follow-up observations, and data reduction. Section~\ref{sec:spec_analysis} describes the spectroscopic evolution within the first $\sim 6$ months after the explosion. Section~\ref{sec:photo_analysis} presents the light curves of SN\,2021ukt and comparisons with the grid of model light curves generated by \texttt{MESA}~\citep{Paxton_2011, Paxton_2013, Paxton_2015, Paxton_2018, Paxton_2019, Jermyn_2023} and \texttt{STELLA}~\citep{Blinnikov_1998, Blinnikov_2000, Blinnikov_2006, Blinnikov_2004, 2005AstL...31..429B}. This model grid was computed as a set of single-star short-plateau (tens of days) light curves that are further investigated by~\cite{2021ApJ...913...55H}. In Section~\ref{sec:disc}, we discuss the interpretations of the progenitor of SN\,2021ukt and compare our results with those of other well-studied SNe. Our discussion and final remarks are given in Section~\ref{sec:summ_conclu}.

\begin{figure*}[t]
    \centering
    \includegraphics[width=\linewidth]{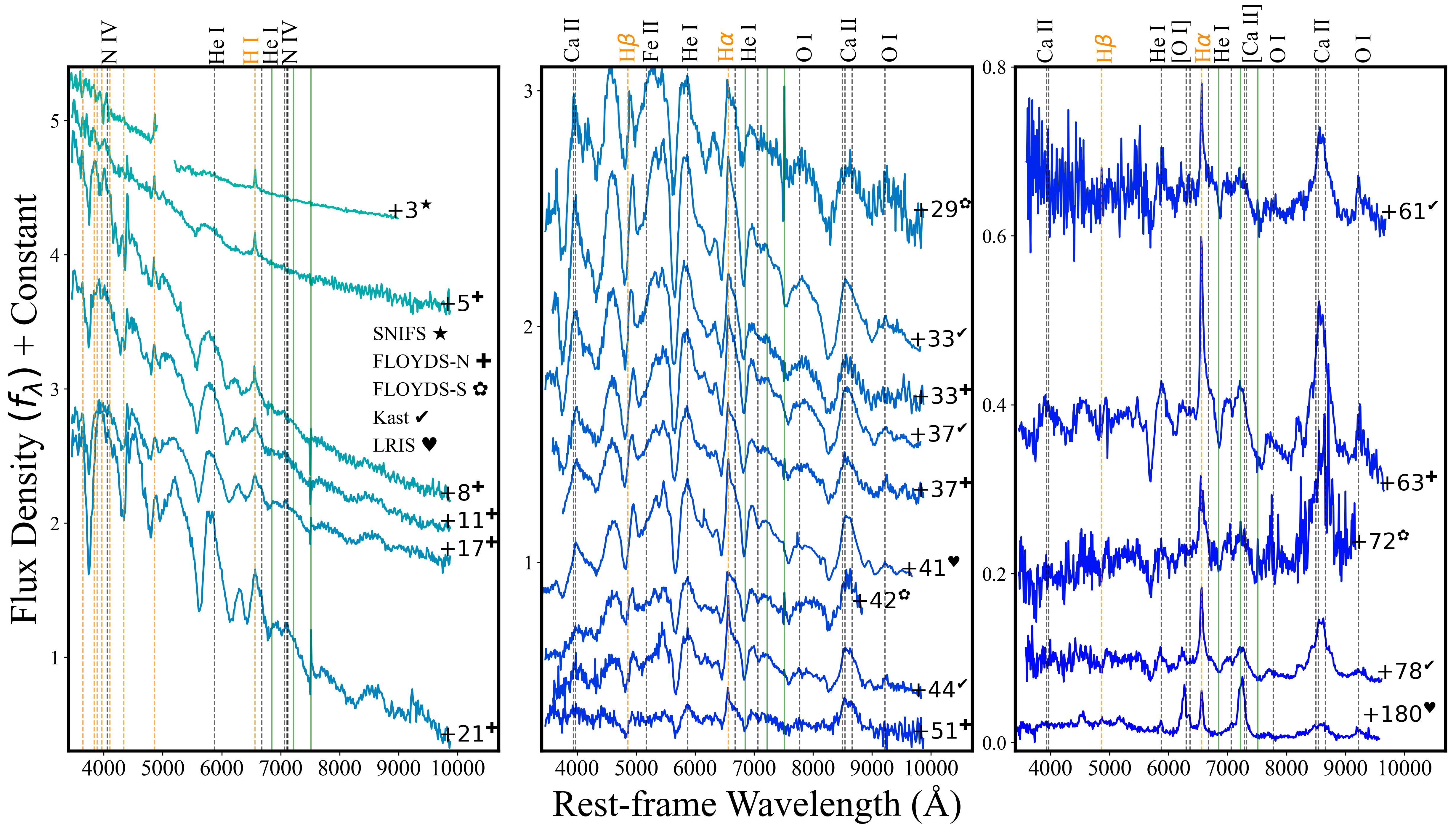}
    \caption{Spectroscopic evolution of SN\,2021ukt. The values on the right of the spectra correspond to the phases between the estimated time of explosion (MJD 59424 $\pm$ 2.5) and the spectroscopic observation date.
    A SNIFS instrumental artifact in the range $\sim 5000$--5200\,\AA\ was manually discarded. Telluric lines in the spectra are labeled with solid green vertical lines. 
    \textit{Left:} The early spectra obtained between days +3 and +11 exhibit narrow, SN\,IIn-like Balmer lines. The \ion{H}{1} emission persists as the ejecta cool and the photosphere recedes. 
    \textit{Middle:} 
    Between days +8 and +51, the spectra manifest broad P~Cygni profiles. H$\alpha$ emission is superimposed on broad \ion{He}{1} $\lambda$6678. \textit{Right:} After $\sim$ day 61, the spectra are dominated by \ion{O}{1}, \ion{Ca}{2}, and H$\alpha$ emission features, which may be attributed to the ejecta interacting with a radially extended CSM profile. 
    SN\,2021ukt appears to be fully nebular by $\sim 6$ months, possibly as early as +78 days.}
    \label{fig:spec}
\end{figure*}

\begin{figure}[h]
    \includegraphics[width=\linewidth]{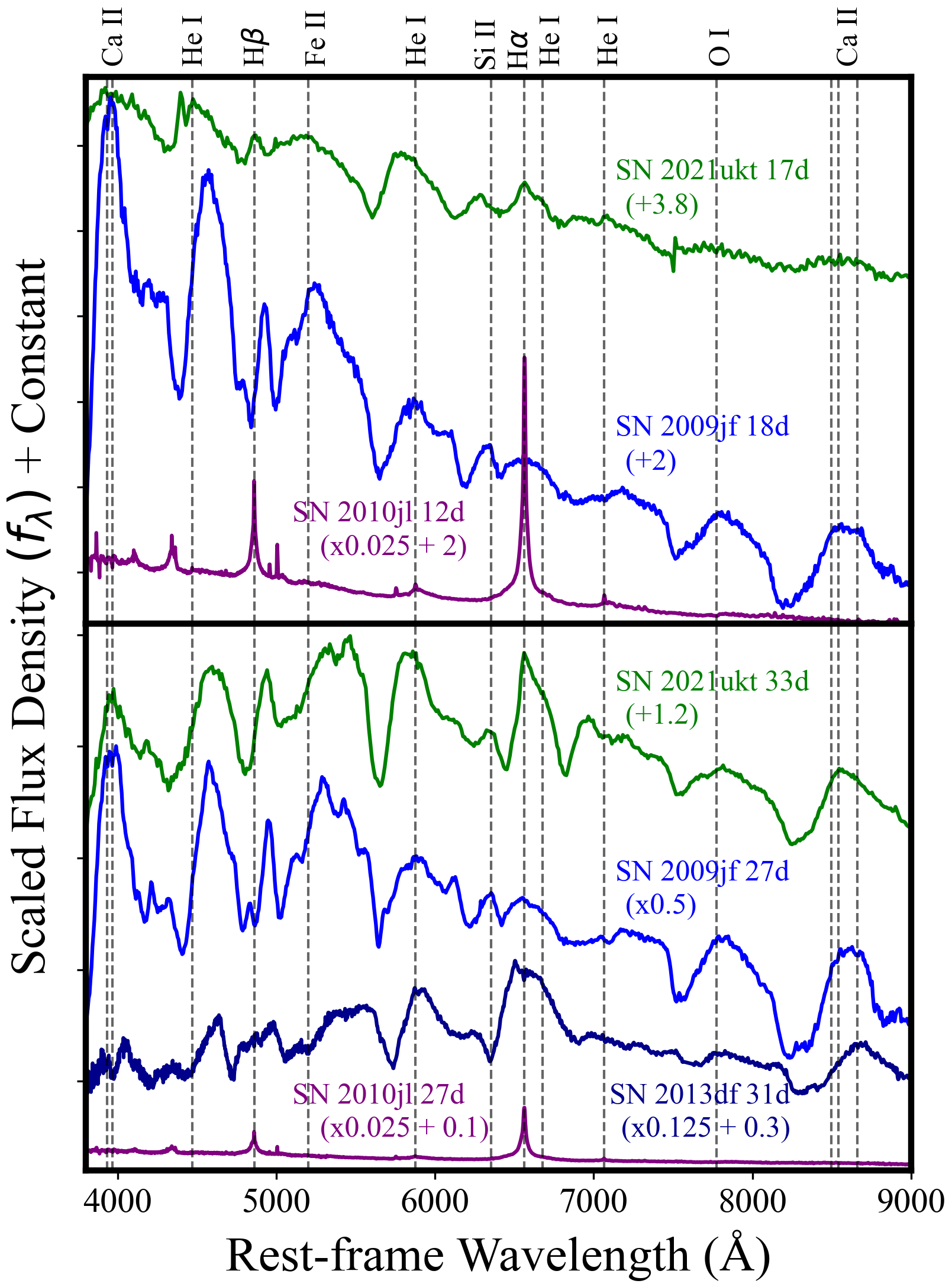}
    \caption{Spectra of SN\,2021ukt (green) at different epochs compared with the spectra of SN\,Ib\,2009jf (blue), SN\,IIb\, 2013df (dark blue), and SN\,IIn 2010jl (purple) at similar phases. Vertical dashed lines mark rest-frame wavelengths of various features. The comparison with SN\,IIb\,2013df is especially striking in the bottom panel, suggesting that SN\,2021ukt was an SN\,IIb at this stage; however, the better-developed \ion{He}{1} lines in SN\,2021ukt, and the bump at the rest wavelength of H$\alpha$ instead possibly being the emission component of the \ion{He}{1} $\lambda$6678 line, argue that it was actually an SN\,Ib as concluded by \citet{yesmin2025spectral}. 
    The spectra of SNe\,2009jf, 2013df, and SN 2010jl were obtained from the Berkeley SuperNova DataBase (SNDB;~\citealt{2012MNRAS.425.1789S};~\citealt{2016MNRAS.461.3057S}). The phases of the spectra of SNe\,2021ukt, 2009jf, and 2013df correspond to the days between first light and the observation date, while the phases of SN\,2010jl correspond to the days between first observation and the observation date. For the purpose of presentation, all spectra were binned to 10\,\AA, and shifted and scaled arbitrarily.
    }\label{fig:spec_comp_ed}
\end{figure}

\begin{figure}[h]
    \includegraphics[width=\linewidth]{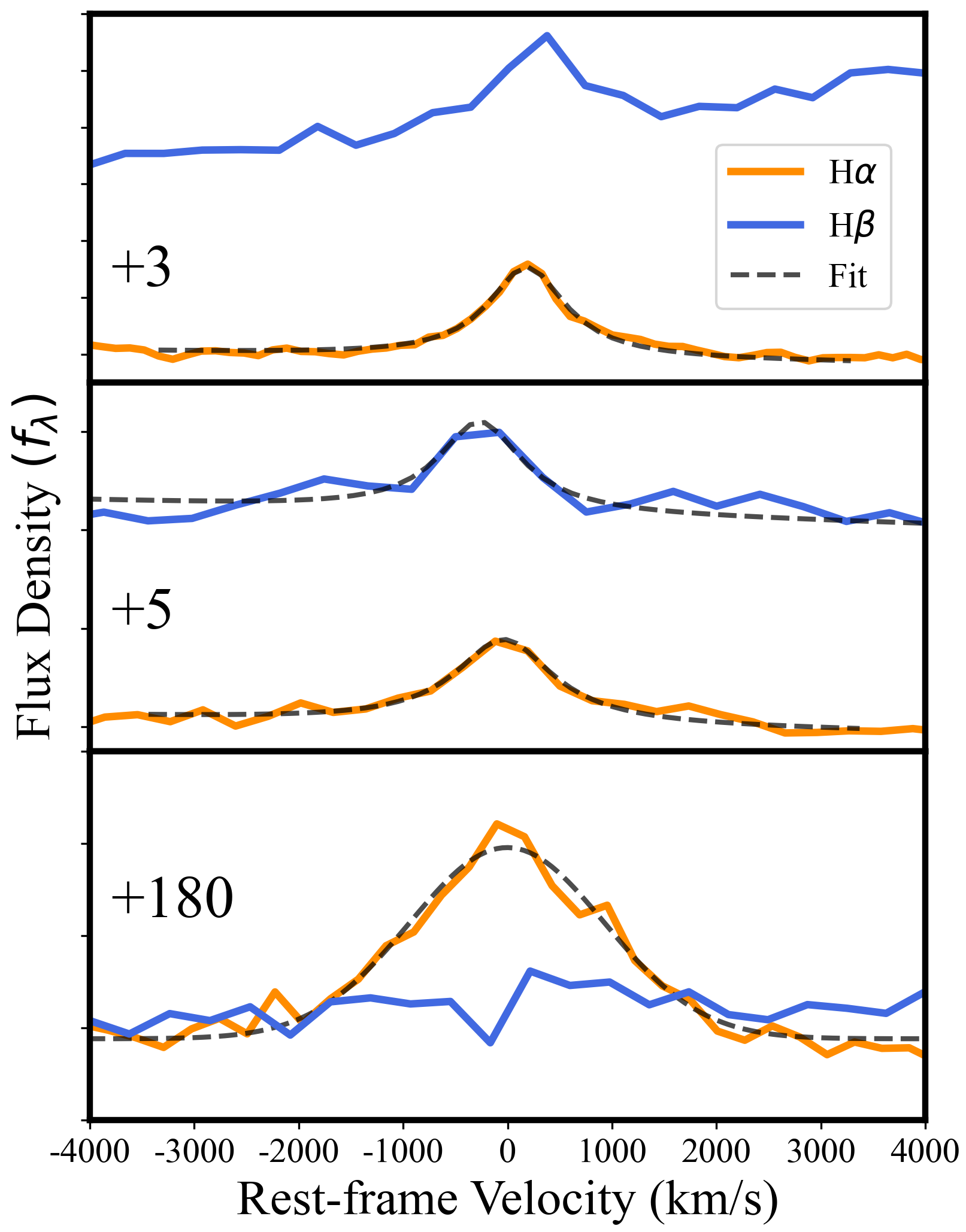}
    \caption{Velocity profiles of the Balmer lines of SN\,2021ukt at days +3, +5, and +180 as presented in the upper, middle, and lower panels, respectively. The FWHM in the first two epochs are measured by fitting a Lorentzian function centered on H$\alpha$ $\lambda$6563 and H$\beta$  $\lambda$4861, while the H$\alpha$ line profile in the nebular spectrum is fitted with a Gaussian function. The H$\alpha$ feature broadens from a FWHM $\approx 900$\,km\,$\text{s}^{-1}$ at day +3 to a FWHM $\approx 2200$\,km\,$\text{s}^{-1}$ at day +180.}
    \label{fig:h_vel}
\end{figure}

\section{Observations}\label{sec:obs}

\subsection{Discovery}
SN\,2021ukt (ZTF21abpxquj) was discovered by the Zwicky Transient Facility (ZTF;~\citealt{2019PASP..131a8002B, 2019PASP..131g8001G}) on July 31, 2021, 10:34:47 (UTC dates are used throughout this paper; MJD 59426.44) in the $r$ band at $16.281 \pm 0.031$ mag~\citep{2021TNSTR2632....1D}. The last nondetection was obtained by the Asteroid Terrestrial-impact Last Alert System (ATLAS;~\citealt{2018PASP..130f4505T}) on July 26, 2021, 13:23:31 (MJD 59421.55) in the $o$ band at 18.670 mag. The five-day gap between the last nondetection and the first detection makes the time of first light difficult to constrain accurately. Here we adopt the midpoint between the last nondetection and first detection, namely MJD $59424 \pm 2.5$, as the estimated date of first light. Throughout this paper, all phases are given relative to the estimated date of first light.

We measure the J2000.0 coordinates of SN\,2021ukt on the images obtained by the University of Hawai'i 2.2\,m telescope to be $\alpha$ = $\text{00}^{\text{hr}}\text{49}^{\text{m}}\text{24}^{\text{s}}.86$ and $\delta = -01^{\circ}45'58".82$ in the spiral galaxy UGC\,505. The NASA/IPAC Extragalactic Database \citep{NED} reports a heliocentric redshift of the host galaxy as $z = 0.012886 \pm 0.000014$ \citep{2005ApJS..160..149S}. The \texttt{Python} package \texttt{astropy.cosmology} with standard $\Lambda$CDM cosmology ($\text{H}_{0} = 72$\,km\,$\text{s}^{-1}\,\text{Mpc}^{-1}$, $\Omega_{M} =0.27$, $\Omega_{\Lambda} =0.73$) was used to calculate the luminosity distance $d_L = 54.20 \ \pm \ 0.060 $\, Mpc and distance modulus $\mu = 33.67 \ \pm 0.0024$\,mag, where the $R_{V}=3.1$ extinction law as the Galactic average was adopted~\citep{1989ApJ...345..245C}. Dust-reddening maps from~\cite{2011ApJ...737..103S} were used to estimate the Milky Way line-of-sight extinction $E_{\text{MW}}(B - V) = 0.0416 \pm 0.0008$\,mag. We identified no prominent absorption doublet of Na{\sc\,I\,D} $\lambda\lambda$5890, 5896 at the redshift of the host galaxy UGC\,505; thus, here we assume that the total extinction is dominated by the Galactic component. The strength of such a feature has been adopted as a tracer of the line-of-sight reddening \citep{1997A&A...318..269M}.

All photometry presented in this work has been corrected for cosmological redshift and extinction as a filter-dependent offset; all spectra are dereddened using the \texttt{dust\_extinction} package~\citep{2023ApJ...950...86G}.

\subsection{Observations and Data Reduction}
The first spectrum of SN\,2021ukt was obtained by the University of Hawai'i 2.2\,m telescope (UH88) on Maunakea with the SuperNova Integral Field Spectrograph~\cite[SNIFS;][]{2002SPIE.4836...61A} +1 day after the first detection; see Table~\ref{table:spec_log}. A description of the SNIFS instrument is given by \cite{2013AJ....146...30K}; details of the SNIFS data-reduction pipeline are outlined by \cite{2001MNRAS.326...23B} and \cite{2006ApJ...650..510A}.

As part of the Global Supernova Project (GSP), the Las Cumbres Observatory (LCO) provided early-time photometry using 1\,m telescopes at several sites (LSC, CPT, COJ, TFN, ELP), which were reduced with the \texttt{lcogtsnpipe} automatic reduction pipeline, described in Appendix B of~\cite{2016MNRAS.459.3939V}. 
This package uses \citet{1992AJ....104..340L} standard fields for $BV$-band Vega magnitude calibrations and Sloan Digital Sky Survey (SDSS) standard-star catalogs~\citep{2002AJ....123.2121S} for $gri$-band AB magnitude calibrations~\citep{1983ApJ...266..713O}. 

Additional photometry of SN\,2021ukt was obtained with the 0.76\,m Katzman Automatic Imaging Telescope (KAIT) as part of the Lick Observatory Supernova Search described by~\cite{2001ASPC..246..121F}, as well as by the Nickel 1\,m telescope at Lick Observatory. Both telescopes obtained $BVRI$-band images. The \texttt{IDL}-based pipeline used for the photometric reductions is outlined by~\cite{2019MNRAS.490.2799D},~\cite{2019MNRAS.490.3882S}, and~\cite{vasylyev2022early}.

The spectral sequence of SN\,2021ukt consists of 20 spectra that span from +5 to +72 days. GSP also provided twelve of the spectra, which were taken by the FLOYDS spectrographs on 2\,m telescopes from Haleakala Observatory (en06) and Siding Spring Observatory (en12). All FLOYDS spectra were extracted, processed, wavelength-calibrated, and flux-calibrated following standard procedures using the FLOYDS pipeline~\citep{2014MNRAS.438L.101V}. Five epochs of optical spectra (days +33 to +78) were obtained using the Kast double spectrograph on the Shane 3\,m telescope at Lick Observatory~\citep{miller1993}. The reduction process is described by~\cite{2012MNRAS.425.1789S} and~\cite{2013MNRAS.436.3614S}. 

Two spectra (days +41 and +180) of SN\,2021ukt were obtained with the Keck I 10\,m telescope on Maunakea using the Low Resolution Imaging Spectrometer (LRIS;~\citealt{1995PASP..107..375O}). A description of the LRIS instrument and the spectral reduction process is given by~\cite{2012MNRAS.425.1789S}. 

Figure~\ref{fig:spec} presents the spectral time series of SN\,2021ukt obtained from days +3 to +180. A log of spectroscopic observations of SN\,2021ukt is provided in Table~\ref{table:spec_log}.

\section{Spectroscopic Analysis}\label{sec:spec_analysis}

\begin{figure*}
    \centering
    \includegraphics[width=\linewidth]{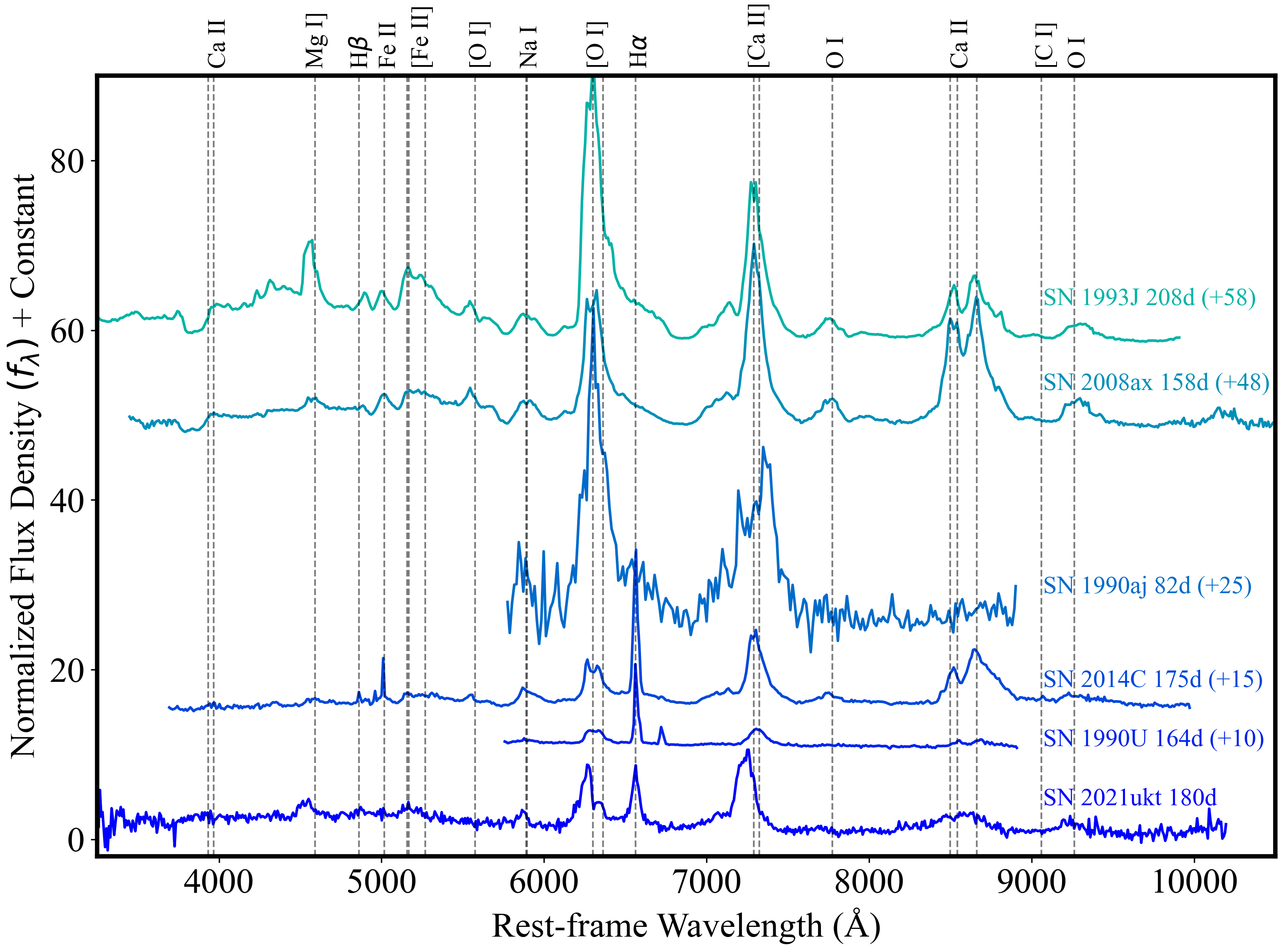}
    \caption{Nebular spectrum of SN\,2021ukt (LRIS, +180 days) compared with the nebular spectra of SNe\,IIb 1993J and 2008ax, SN\,Ib\,1990U, SN\,Ib/Ic 1990aj, and SN\,Ib--IIn 2014C obtained at similar phases. For the purpose of presentation, the underlying continuum of each spectrum was fitted by a low-order polynomial and divided. All presented spectra were arbitrarily shifted. The phases correspond to the time between first light and the observation date, except for SNe\,1990aj and 1990U, which are given relative to the time of the first detection.
    All spectra presented in this figure were sourced from the Berkeley SuperNova DataBase (SNDB;~\citealt{2012MNRAS.425.1789S};~\citealt{2016MNRAS.461.3057S}) and~\cite{2019MNRAS.482.1545S}.} 
    \label{fig:neb_spec}
\end{figure*}

\begin{figure}[h]
    \includegraphics[width=\linewidth]{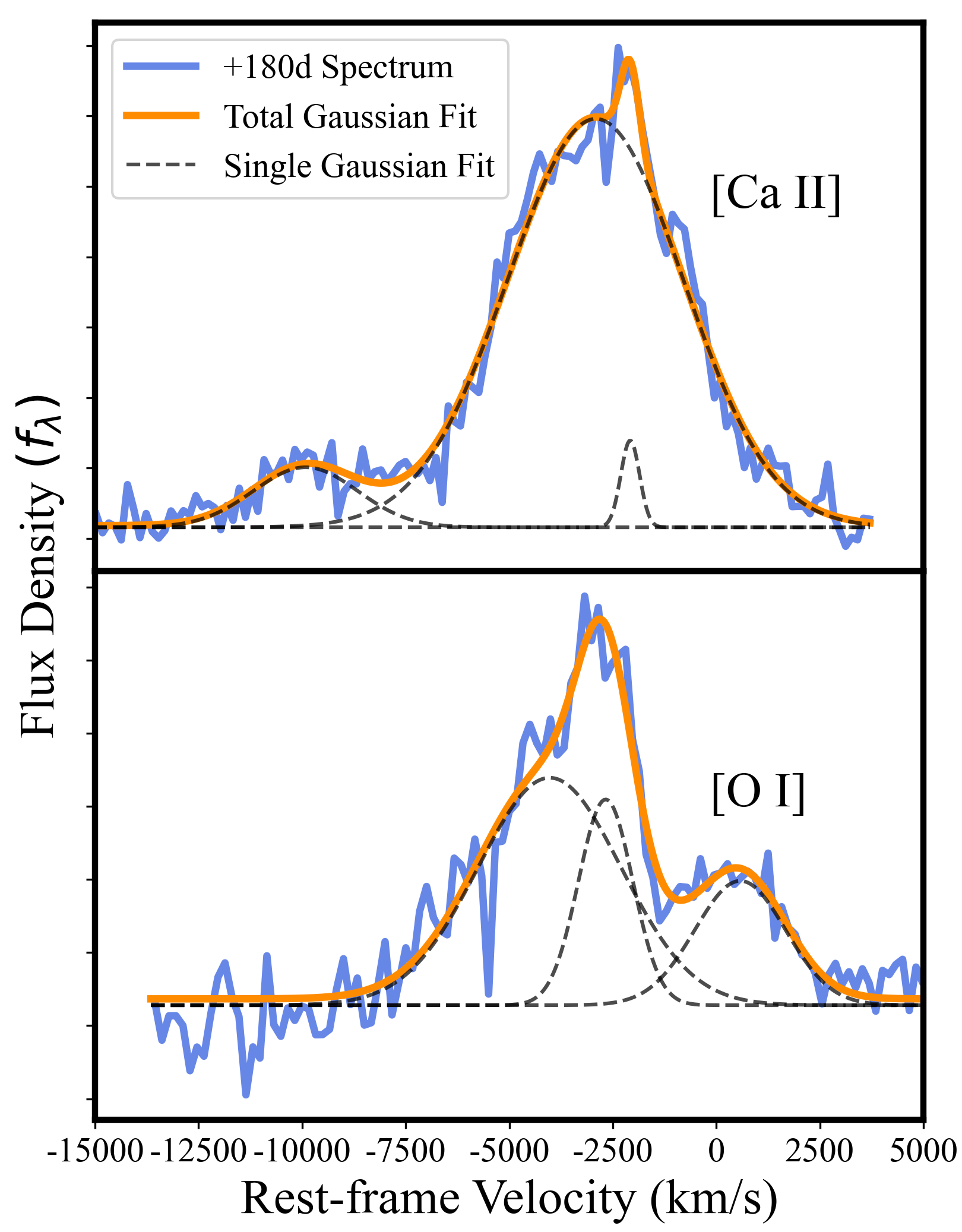}
    \caption{Velocity profiles of emission lines [\ion{Ca}{2}] $\lambda \lambda$7291, 7324 and [\ion{O}{1}] $\lambda \lambda$6300, 6364 at day +180. We first estimate the underlying continuum of the lines of interest by linearly interpolating the flux spectrum from adjacent wavelength region and subtracting it from the observed flux. We then integrate the flux below the best fit to the line profile after adopting a multicomponent Gaussian function to fit the emission features and account for noise. The velocities are estimated by with the rest-frame wavelength of the the midpoint of the doublets.}
    \label{fig:ca_o_ratio}
\end{figure}
We divide the spectral evolution of SN\,2021ukt into three phases, namely the CSM-dominated, ejecta-dominated, and nebular phases. The first spectrum of SN\,2021ukt obtained on day +3 reveals conspicuous, relatively narrow emission lines from ionized CSM (see left panel of Figure~\ref{fig:spec}). Some SN spectra exhibit a quasifeatureless continuum, which can occur when the photosphere is in the region between the reverse and forward shocks. This is known as the cold-dense-shell-dominated phase~\citep{2007ApJ...662.1136C}. Well-observed examples include SN\,1998S~\citep{2016MNRAS.458.2094D} and SN\,2023ixf~\citep{2024Natur.627..759Z, 2025ApJ...988...61Z}. Since these narrow emission lines in the spectrum of SN\,2021ukt persist for the first 11 days, we will call this the CSM-dominated phase. 

After $\sim 3$ weeks, the spectrum is characterized by broad P~Cygni profiles, indicating that the receding photosphere has passed the inner boundary of the CSM and enters the expanding ejecta. This will be called the ejecta-dominated phase. As displayed in Figure~\ref{fig:spec_comp_ed}, the
spectra of SN\,2021ukt at this phase show a remarkable resemblance to those of SNe\,Ib and IIb \citep{1988AJ.....96.1941F,1993ApJ...415L.103F}. 
The spectrum of SN\,2021ukt in the bottom panel of Figure~\ref{fig:spec_comp_ed} seems especially similar to that of SN\,IIb\,2013df; however, the \ion{He}{1} lines in SN\,2021ukt are significantly more well-developed than those in SN\,IIb\,2013df, and  the bump at the rest-frame wavelength of H$\alpha$ in SN\,2021ukt might be caused mostly by the emission component of the P~Cygni profile of \ion{He}{1} $\lambda$6678 instead of by H$\alpha$. Thus, the true classification of SN\,2021ukt may be closer to SN\,Ib than to SN\,IIb, as was also found by \citet{yesmin2025spectral}.

After $\sim 2.5$ months, the ejecta become optically thin and the nebular-phase spectrum is dominated by several prominent emission lines, superimposed on a continuum due to significantly blended emission profiles from various iron-group elements. We will denote this evolutionary stage as the nebular phase of SN\,2021ukt. 

\subsection{CSM-Dominated Phase}\label{sec:spec_phot}
The left panel of Figure~\ref{fig:spec} shows the CSM-dominated spectrum of SN\,2021ukt (+3 days, SNIFS), which is characterized by various emission lines superimposed on a blackbody continuum of $\sim 10,000$\,K~\citep{2021TNSCR2645....1H}. The strongest emission lines can be identified as the characteristic \ion{H}{1} Balmer series (in blue), which are often presented in the early spectra of SNe\,IIn, such as SN\,2010jl~\citep{2014ApJ...781...42O} and SN\,2006gy~\citep{smith2010spectral}. The emission lines arise from the corresponding ionic species within the ambient CSM photoionized by the initial shock wave as it runs through and manifest in the spectrum as broad, Lorentzian wings. The broad wings of the features are most likely due to scattering by free electrons in the unshocked, photoionized CSM~\citep{2001MNRAS.326.1448C, 2009MNRAS.394...21D, 2018MNRAS.475.1261H}.

To measure the width of such line profiles, after subtracting the underlying continuum through an arbitrary linear interpolation, we fit a Lorentzian function to the emission profiles. Figure~\ref{fig:h_vel} portrays the H$\alpha$ profile with a FHWM = $890 \pm 40$\,km\,$\text{s}^{-1}$ at day +3 and FHWM = $1130 \pm 90$ \,km\,$\text{s}^{-1}$ at day +5.

Weak emission lines include \ion{He}{1} $\lambda$5876, \ion{N}{4} $\lambda$4058, and \ion{N}{4} $\lambda\lambda$7109, 7123. These lines persisted for the first $\sim$ eight days, where \ion{He}{1} $\lambda$6678 and $\lambda$7065 also appear. The growing strength of the broad \ion{He}{1} $\lambda$5876 and \ion{Ca}{2} H\&K $\lambda \lambda$3933, 3968 absorption lines indicates the evolution to the ejecta-dominated phase.

\subsection{Ejecta-Dominated Phase}\label{sec:spec_ejec}
On day +5, the photosphere has already receded into the fast-moving ejecta, as evidenced by the emergence  of the broad P~Cygni profiles labeled in the middle panel of Figure~\ref{fig:spec}. The prompt emergence of the helium-rich layer of the ejecta as indicated by the development of the P~Cygni profile of the \ion{He}{1}\,$\lambda$5876 line indicates a close resemblance to that observed in other SNe\,Ib and IIb at similar phases (see Figure~\ref{fig:spec_comp_ed}; also, e.g., \citealp{valenti2011sn}). The evolution of SN\,2021ukt from its CSM-dominated phase to its ejecta-dominated phase marks an unprecedented transition from its original SN\,IIn classification to an SN\,Ib (possibly SN\,IIb).

As illustrated in Figure~\ref{fig:spec}, the P~Cygni profile of the \ion{He}{1}\,$\lambda$5876 line is fully developed. We estimate the velocity of the outer He-rich layer of the ejecta by fitting a Gaussian distribution using the \texttt{scipy.optimize} package to the absorption minimum of this feature. The absorption minimum is measured at a velocity of $\sim 15,000$\,km\,$\text{s}^{-1}$, consistent with that measured from the absorption minima of the \ion{He}{1} $\lambda$6678 and $\lambda$7065 lines. At +33 days, the Kast and FLOYDS spectral continua are consistent with a blackbody of $\sim 5500$\,K, and the \ion{He}{1} $\lambda$5876 absorption minimum shifts to $\sim 11,000$\,km\,$\text{s}^{-1}$. At +63 days, continuum emission ceases and the \ion{He}{1} $\lambda$5876 absorption minimum further shifts to $\sim 9000$\,km\,$\text{s}^{-1}$. These values are consistent with the velocities measured by \cite{yesmin2025spectral}. 

\ion{He}{1} $\lambda$5876 exhibits a flat-topped emission profile in Figure~\ref{fig:spec} from +11 to +44 days, while the emission peak of \ion{He}{1} $\lambda$7065 is relatively narrow. From +8 to +72 days, an intermediate-width H$\alpha$ component is superimposed on the broad emission of \ion{He}{1} $\lambda$6678 and resembles a ``shark-fin'' shape. At +44 days, the strength of \ion{He}{1} $\lambda$6678 weakens relative to the H$\alpha$ component, highlighting the persistence of H$\alpha$ emission. This suggests that either there is contamination from a foreground \ion{H}{2} region or there is ongoing interaction with slow-moving H-rich CSM. H$\alpha$ emission persists from the CSM-dominated phase, the ejecta-dominated phase, and the nebular phase.

\subsection{Nebular Phase}\label{sec:spec_nebu}
The right panel of Figure~\ref{fig:spec} shows the spectral evolution of SN\,2021ukt from +63 days to +180 days. In the +63 day FLOYDS spectrum, we observe weaker emission and shallower absorption profiles of lines that previously dominated the spectrum, such as \ion{He}{1}, \ion{Ca}{2} H\&K, and \ion{Fe}{2}. This coincides with the growing emission of the \ion{Ca}{2} near-infrared (NIR) triplet alongside forbidden lines [\ion{O}{1}] $\lambda \lambda$6300, 6364 and [\ion{Ca}{2}] $\lambda \lambda$7291, 7324. From +63 to +78 days, the \ion{He}{1} $\lambda$6678 line weakens, leaving only H$\alpha$ emission along with the forbidden lines.

SN\,2021ukt was also observed by Keck/LRIS on day +180, when the SN had entered the nebular phase (see the right panel of Figure~\ref{fig:spec} and Figure~\ref{fig:neb_spec}). The day +180 spectrum of SN\,2021ukt exhibits several characteristic prominent emission lines, including [\ion{O}{1}]\,$\lambda \lambda$6300, 6364, [\ion{Ca}{2}]\,$\lambda \lambda$\,7291, 7324, \ion{Ca}{2} NIR triplet, and H$\alpha$. Furthermore, \ion{Mg}{1}] $\lambda$4571, \ion{Na}{1} $\lambda$5890, and \ion{O}{1} $\lambda$9221 are tentatively detected. The emission peaks of [\ion{O}{1}] $\lambda \lambda$6300, 6364 and [\ion{Ca}{2}] $\lambda \lambda$7291, 7324 also exhibit a $\sim 1500$\,km\,s$^{-1}$ blueshift with respect to the rest frame. 

Oxygen is one of the products of a series of hydrostatic burning phases that exhibits prominent features in nebular-phase spectra of CCSNe. The oxygen content in the SN ejecta increases
with the zero-age sequence-mass (ZAMS) of the progenitor star. As fuel for calcium production during the explosive burning phase, the strength of [\ion{O}{1}]\,$\lambda \lambda$6300,\,6364 and the [\ion{Ca}{2}]/[\ion{O}{1}] flux ratio provide sensitive indicators of the ZAMS mass of the progenitor of CCSNe~\citep{1989ApJ...343..323F, 2012A&A...546A..28J, 2014MNRAS.439.3694J, 2015A&A...579A..95K, 2021A&A...656A..61D, 2022ApJ...928..151F}.

We estimate the ZAMS mass by measuring the [\ion{Ca}{2}]/[\ion{O}{1}] flux ratio (see the caption of Figure \ref{fig:ca_o_ratio}).
From the +180 day spectrum, we measure [\ion{Ca}{2}]/[\ion{O}{1}] = 1.4\,$\pm$\,0.05. Some SNe with greater oxygen-core masses (i.e., [\ion{Ca}{2}]/[\ion{O}{1}] $< 1$) represent a population of SN progenitors in the $M_{\text{ZAMS}}$ range of 20--40\,$\text{M}_{\odot}$ (e.g., SN\,Ib\,2009jf,~\citealt{valenti2011sn}; SN\,Ic 1998bw,~\citealt{mazzali2001nebular}; SN\,Ic 2008ap,~\citealt{mazzali2002type}). Considering the relatively low oxygen-core mass of SN\,2021ukt, this suggests an upper limit of $M_{\text{ZAMS}}\lesssim\text{20}\,\text{M}_{\odot}$.

Following the methods outlined by~\cite{jerkstrand2015late}, we estimate the ZAMS mass of SN\,2021ukt by comparing the [\ion{O}{1}] $\lambda \lambda$6300, 6364 luminosity normalized to the $^{56}\text{Co}$ decay power of SNe\,IIb (see Eq. 1 of~\citealt{jerkstrand2015late}). Using the methods for measuring the flux described above, we convert the  [\ion{O}{1}] flux to luminosity with our distance measurement and assuming a $^{56}\text{Ni}$ mass of 0.04\,$\text{M}_{\odot}$ (see Section~\ref{sec:model_results}). Therefore, we calculate $L_{\text{norm}}(180)$ = 0.004\,$\pm$\,0.0001 and suggest that this value plausibly matches models 12A and 12D with $M_{\text{ZAMS}} = 12\,\text{M}_{\odot}$ (see Figure 15 of~\citealt{jerkstrand2015late}). This ZAMS mass is consistent with our discussion above.

\begin{figure*}[t]
    \centering
    \includegraphics[width=\linewidth]{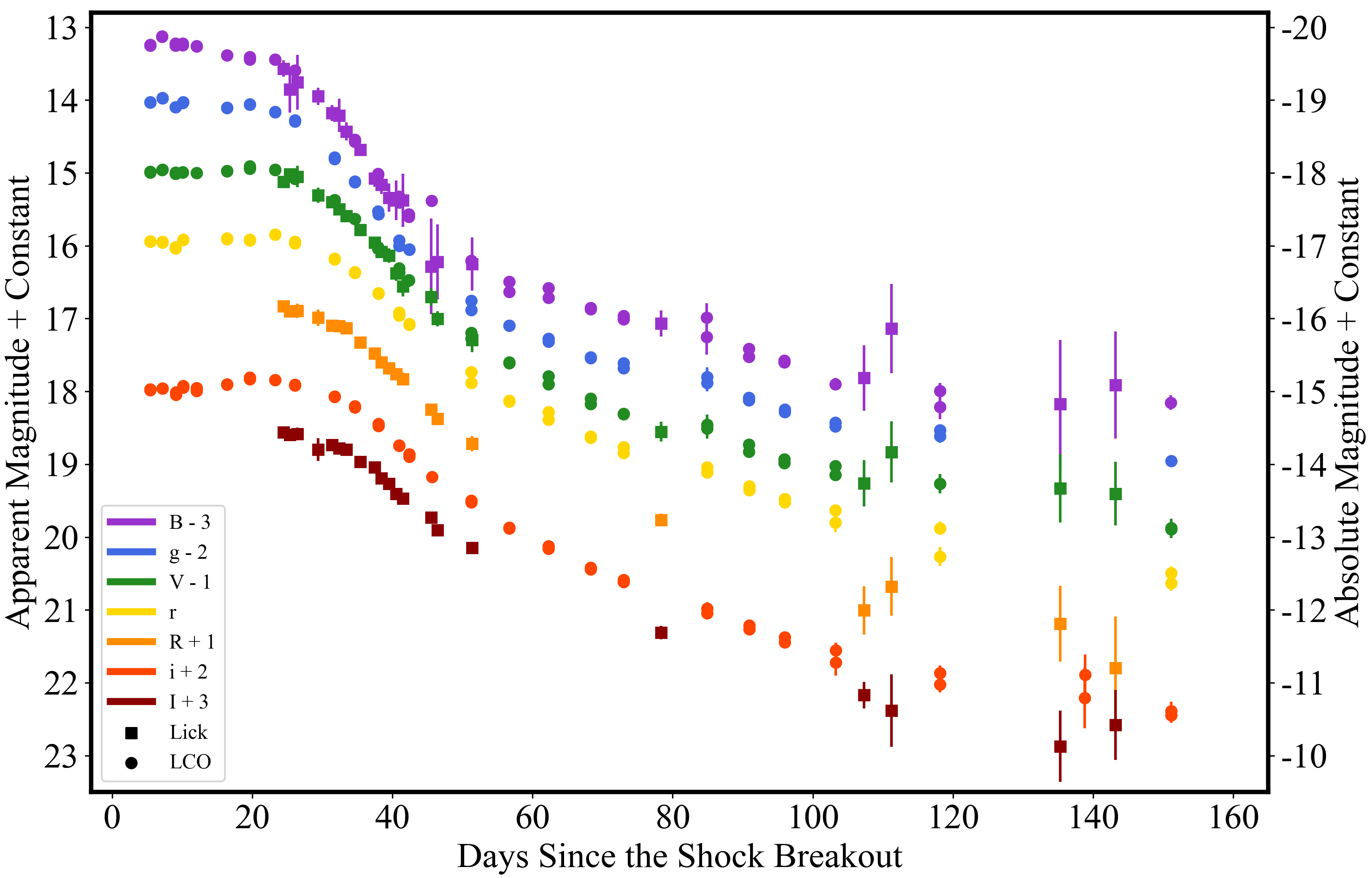}
    \caption{$BgVri$-band light curves of SN\,2021ukt. All phases are given relative to the estimated time of the first light at MJD $59424 \pm 2.5$ and are corrected to the rest frame. All photometry has been corrected for the Galactic reddening (see Section~\ref{sec:obs}).}
    \label{fig:lc}
\end{figure*}

\section{Photometric Analysis}\label{sec:photo_analysis}

\subsection{Optical Light Curve}
Figure~\ref{fig:lc} shows the Lick + LCO optical light curves of SN\,2021ukt during the first five months after the first light. SN\,2021ukt peaked at $V\approx14$\,mag and maintained a nearly constant luminosity for a period of $\sim 25$ days past maximum. During such an optical plateau phase, a subtle rise in the $R$ light curve begins around 15 days and peaks at 20 days. We then observe a gradual decrease in flux before the light curve declines to the $^{56}\text{Co}$-powered radioactive tail, which is typical of other SNe\,II.

The plateau observed in SN\,2021ukt may be attributed to an optically thick, He-rich envelope that undergoes interaction with an H-rich wind. The ejecta-CSM interaction can produce excess emission, which may lead to a higher peak luminosity and extend the duration of the plateau phase (see Figure 8 of~\citealt{2021ApJ...913...55H}; also, see~\citealt{2025ApJ...978...56M}). However, CSM interaction alone is not enough to be responsible for extending the plateau. Additional factors such as the progenitor radius and explosion energy may also affect the length of the plateau in CCSNe light curves \citep{2019ApJ...879....3G}. Given the limited number of observations of short-plateau SNe, SN\,2021ukt represents a unique intermediate case within the continuum of SN\,IIL and SN\,IIP light curves~\citep{2014ApJ...786...67A}. 

\subsection{Light-Curve Modeling}\label{sec:model}
In this section, we compare the optical light curves of SN\,2021ukt with grids of model light curves generated with \texttt{MESA}+\texttt{STELLA}. \texttt{MESA} is a stellar hydrodynamical code that simulates the evolution of a star from its pre-main sequence to iron core collapse, followed by the injection of thermal energy that leads to shock breakout. The properties of this evolved star are then used as input for \texttt{STELLA}, a one-dimensional (1D) multigroup radiative-transfer code that simulates how light interacts with the ejected stellar material from shock breakout to the radioactive tail. \texttt{MESA}+\texttt{STELLA} compute the pseudobolometric ($UBVRI$) light curves used for our analysis of SN\,2021ukt. The pseudobolometric ($BVri$) light curve of SN\,2021ukt is constructed using the \texttt{Python}-based code \texttt{SuperBol}, which corrects for cosmological redshift and extinction~\citep{Nicholl_2018}.

\begin{figure*}[t]
    \centering
    \includegraphics[width=\linewidth]{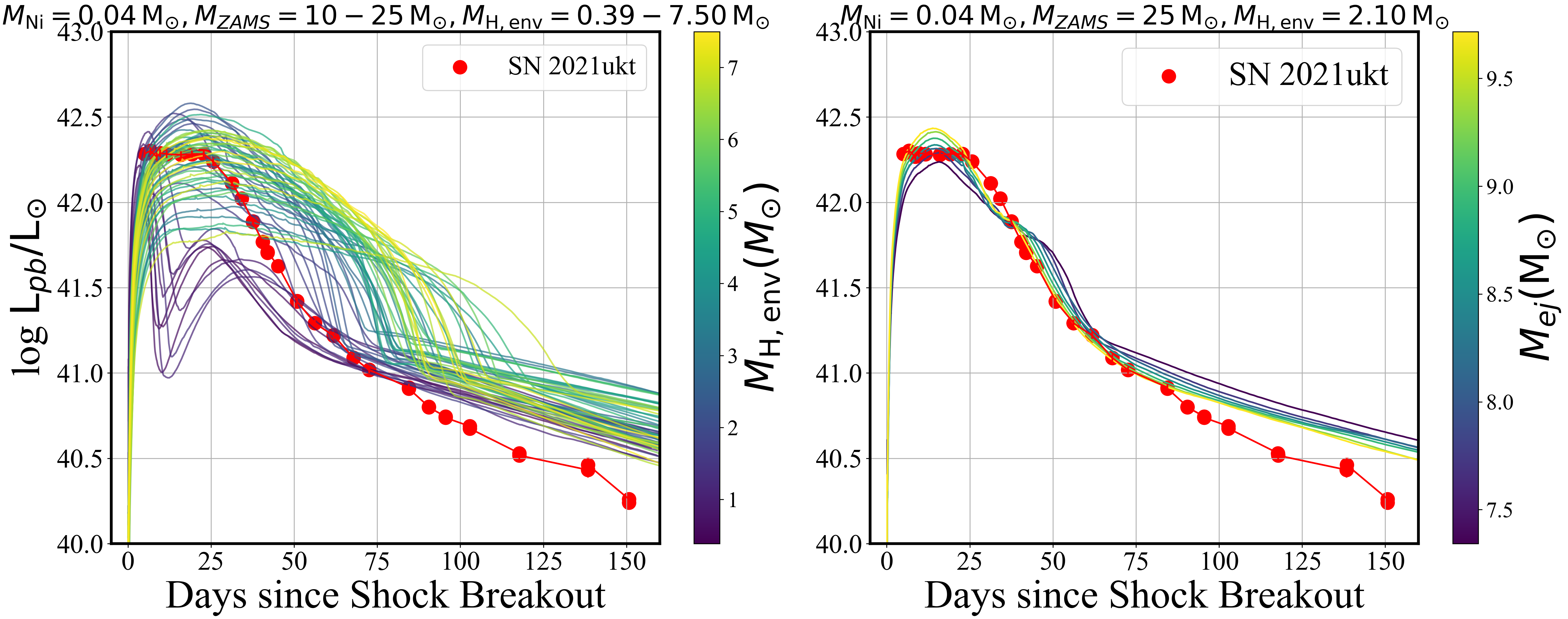}
    \caption{\texttt{MESA}+\texttt{STELLA} light-curve model grid obtained from~\citet{2021ApJ...913...55H}. The model grid's pseudobolometric luminosity ($L_{\rm pb}$, erg\,$\text{s}^{-1}$) integrates the $UBVRI$ bands, while the $L_{\rm pb}$ (erg\,$\text{s}^{-1}$) of SN\,2021ukt integrates the $BVri$ bands. We constrained $M_{\text{Ni}} = 0.04\,\text{M}_{\odot}$ --- the minimum computed $M_{\text{Ni}}$ value --- since it best correlated with the radioactive tail of SN\,2021ukt. \textit{Left:} We compare SN\,2021ukt to the model grid by varying the H-rich envelope mass ($M_{\rm H,env}$), where more-massive $M_{\rm H,env}$ is represented by lighter colors. SNe\,IIb tend to exhibit double-peaked light curves, which are observed in the darker, lower-$M_{\rm H,env}$ light curves. As $M_{\rm H,env}$ is increased, we observe the transition from short-plateau $\rightarrow$ SNe\,IIL $\rightarrow$ SNe\,IIP. \textit{Right:} The best-correlated set of synthetic light curves, configured to  $M_{\text{ZAMS}} = 25\,\text{M}_{\odot}$ and $M_{H,env} = 2.1\,\text{M}_{\odot}$. We further analyze this configuration identifying the ejecta masses ($M_{\rm ej}$) of each individual synthetic light curve; more-massive $M_{\rm ej}$ tend to correlate with greater explosion energies, which produce light curves with brighter peaks and dimmer tails.}\label{fig:models}
\end{figure*}

\begin{figure*}
    \centering
    \includegraphics[width=\linewidth]{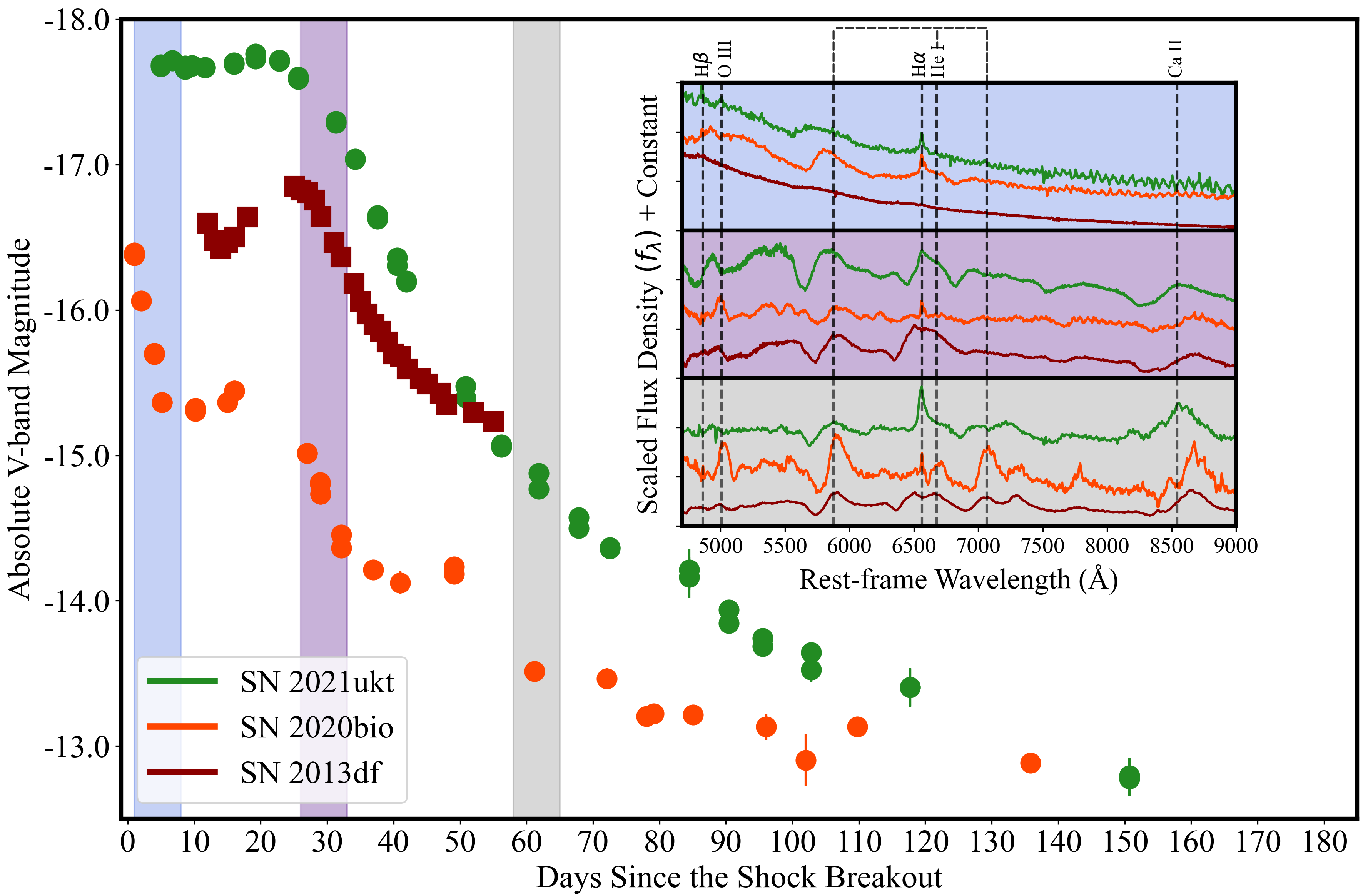}
    \caption{The $v$-band light curve of SN\,2021ukt (green) alongside spectra obtained during the three characteristic phases as described in Section~\ref{sec:spec_analysis}. Light curves and spectral evolution of SNe\,IIb 2020bio (orange) and 2013df (dark red) are also displayed for comparison. The similar phases of the SNe are represented in as the highlighted regions on the light curve. All photometry is corrected for the Galactic reddening.}
    \label{fig:line_evolution}
\end{figure*}

\subsubsection{Model Configuration}\label{sec:model_config}

\begin{table}[ht]
    \centering
    \caption{\texttt{MESA}+\texttt{STELLA} Model Grid Parameters}
    \begin{tabular}{lll}
        \hline
        \textbf{Parameter} & \vline \ \textbf{Value/Range} & \vline \ \textbf{Increment} \\
        \hline
        $M_{\text{ZAMS}}\, (\text{M}_{\odot})$ & \vline \ 10.0--25.0 & \vline \ 2.5  \\
        \hline
        $Z$ ($\text{Z}_{\odot}$) & \vline \ 0.3 & \vline \ fixed \\
        \hline
        $\eta_{\text{wind}}$ & \vline \ 0.0--3.0 & \vline \ 0.1 \\
        \hline
        $\nu/\nu_{\text{crit}}$ & \vline \ 0.0 & \vline \ fixed \\
        \hline
        $E_{\text{exp}}$ ($10^{51}$\, erg) &  \vline \ 0.4--2.0 & \vline \ 0.2 \\
        \hline
        $M_{\text{Ni}} \, (\text{M}_{\odot})$  &  \vline \ 0.04--0.1 & \vline \ 0.03 \\
        \hline
        \label{table:models}
    \end{tabular}
\end{table}

The effect of H-rich envelope stripping on SN\,II light curves was explored by~\citet{2021ApJ...913...55H}. Light-curve models ranging from large to small masses of the H-rich envelope retained by the exploding progenitor star have been produced to account for the observed luminosity evolution of the SNe\,IIP-IIL-short plateau-IIb sequence, respectively (see Figure 5 of~\citealt{2021ApJ...913...55H}). The authors explore how various SN progenitor models can produce short-plateau light curves, which we use to constrain the progenitor of SN\,2021ukt.
Table~\ref{table:models} summarizes the parameters used in the \texttt{MESA}+\texttt{STELLA} light-curve model grid for a single-star progenitor. Further details of these models and methods are described in the Appendix of~\citet{2021ApJ...913...55H}. The model grid is computed using a wide range of $M_{\text{ZAMS}}$, progenitor wind mass-loss scaling factors ($\eta_{\text{wind}}$), explosion energies ($E_{\text{exp}}$) and $^{56}\text{Ni}$ masses ($M_{\text{Ni}}$). All models in the grid are computed for a nonrotating progenitor with a fixed metallicity of $0.3\,\text{M}_{\odot}$.

\subsubsection{Results}\label{sec:model_results}
The $\sim 25$ day light-curve plateau of SN\,2021ukt makes this object an ideal candidate for comparison with short-plateau models. Figure~\ref{fig:models} highlights a set of \texttt{MESA}+\texttt{STELLA} synthetic light curves computed with varying hydrogen envelope masses $M_{\rm H,env}$. Higher values of $M_{\rm H,env}$ are represented with progressively lighter colors. The range of $1\,{\rm M}_{\odot}\lesssim M_{\rm H,env} \lesssim 2\,{\rm M}_{\odot}$ distinguishes the short-plateau light curves, which we will discuss further below. Figure A1 of~\cite{2021ApJ...913...55H} partitions four ranges of $M_{\rm H,env}$ to more clearly highlight the photometric diversity, where the authors identify the range of $2\,{\rm M}_{\odot}\lesssim M_{\rm H,env} \lesssim 15\,{\rm M}_{\odot}$ as the SN\,IIL/IIP regime.

The plateau phase was not well reproduced by any of the grid models. The right panel of Figure~\ref{fig:models} compares a set of synthetic light curves that best correlate with the photospheric phase of SN\,2021ukt. The light curve of SN\,2021ukt is best matched with a 25\,M$_{\odot}$ progenitor with $M_{\text{Ni}} =0.04\,{\rm M}_{\odot}$ and $M_{\rm H,env} = 2.10\,{\rm M}_{\odot}$. The progenitors of short-plateau SNe in~\cite{2021ApJ...913...55H} were found to have ZAMS masses of 18--22\,M$_{\odot}$, which implies that SN\,2021ukt belongs to a higher ZAMS mass range of short-plateau SN progenitors. Higher values of $E_{\text{exp}}$ result in larger $M_{\rm ej}$, which produce light curves with brighter peaks and dimmer radioactive tails. Since the radioactive tail of SN\,2021ukt is fainter and steeper than that of the synthetic light curves, this suggests that SN\,2021ukt either has a low-$M_{\text{Ni}}$ (upper limit $\sim 0.04\,{\rm M}_{\odot}$), or a low-density, high-velocity ejecta with inefficient trapping, or a combination of both conditions.

\section{Discussion and Comparisons}\label{sec:disc}

\begin{figure*}
    \centering
    \includegraphics[width=\linewidth]{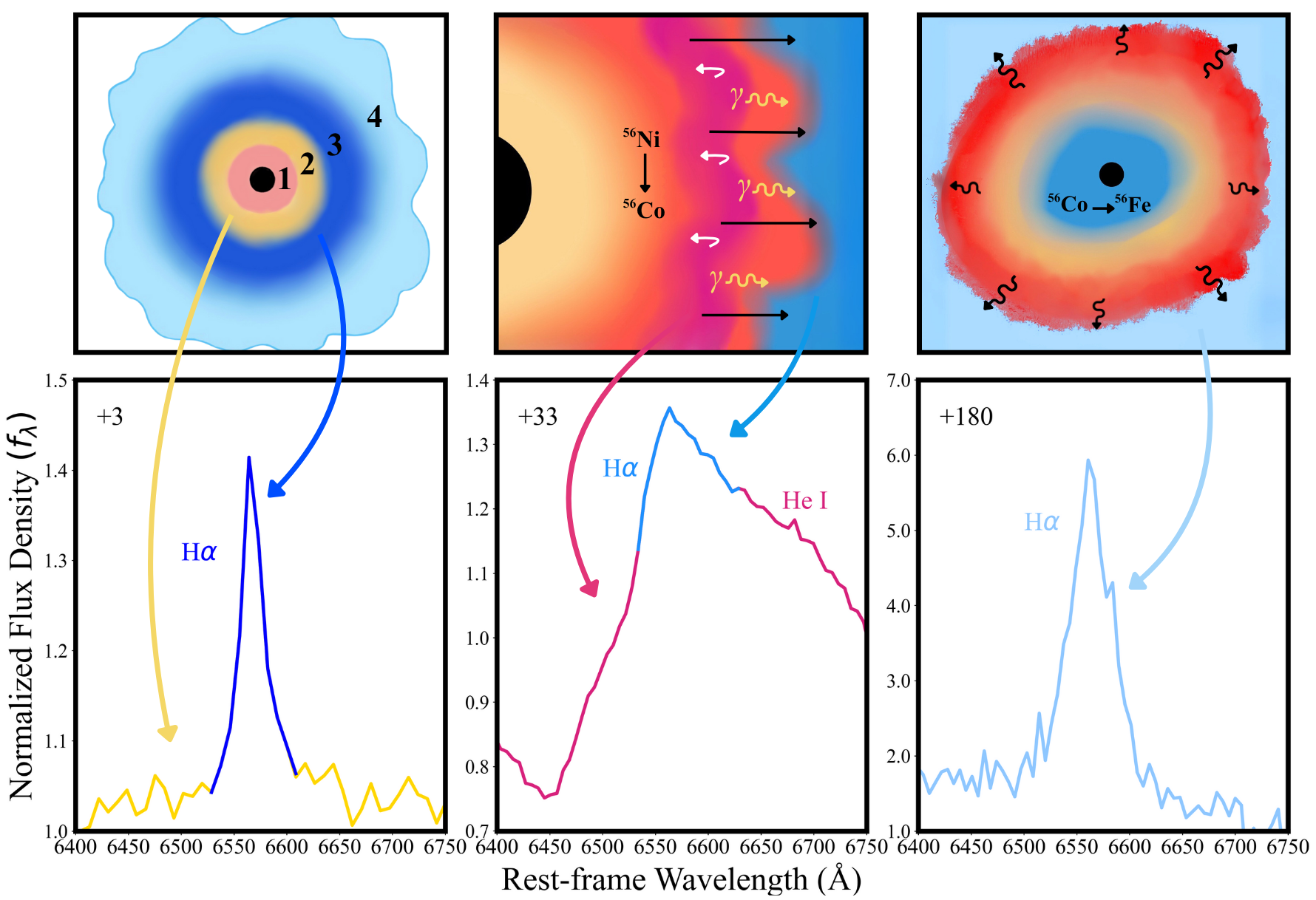}
    \caption{Schematic interpretation of the temporal evolution of the spectrophotometric properties of SN\,2021ukt. Numbers in the top-left 
    panel label the following: \textbf{1}, the helium-rich ejecta; \textbf{2}, the approximate location of the photosphere at day 3; \textbf{3}, and the radially confined CSM; and \textbf{4}, the extended shell of CSM. 
    \textit{Left:} The spectrum at day +3 is dominated by the photosphere located near the shocked region between the ejecta and the photoionized CSM, producing emission lines superposed on a blue continuum. 
    \textit{Middle:} At day +33, the nearby CSM is swept up and outer layers of the He-rich ejecta become optically thin. Emission from the ejecta continue to ionize the extended, H-rich shell, producing a ``shark-fin'' line profile with broad \ion{He}{1} and intermediate-width H$\alpha$. 
    \textit{Right:} At day +180, the inner layers of the ejecta become optically thin and shock interaction with the extended shell produces H$\alpha$ emission.}
    \label{fig:aud_draw}
\end{figure*}

Figure~\ref{fig:line_evolution} compares SN\,2021ukt with SNe\,IIb 2013df and 2020bio. Similarly to SN\,2021ukt, the first spectra of SNe\,2013df and 2020bio obtained within $\lesssim 5$ days after the SN explosion are dominated by \ion{H}{1} emission that subsequently transitions to show strong \ion{He}{1} emission~\citep{10.1093/mnras/stu1837,2023ApJ...954...35P}. SN\,2013df revealed evidence of interaction through the presence of H$\alpha$ in the late-time spectrum \citep{2014MNRAS.445.1647M}, while the early interaction of SN\,2020bio is evidenced by a remarkably blue early-time light curve~\citep{2023ApJ...954...35P}. This could suggest that the progenitors of SNe\,2013df and 2020bio had been stripped of their outer envelopes prior to the explosion. One natural explanation for stripping is offered by binarity, such as SNe\,1993J~\citep{1993ApJ...415L.103F,1994ApJ...429..300W,maund2004massive,stancliffe2009modelling}, 2011dh~\citep{benvenuto2012binary,2013MNRAS.436.3614S}, 2013df~\citep{2014AJ....147...37V}, and 2020bio~\citep{2023ApJ...954...35P}. Although SN\,2021ukt did not show remarkably blue light curves in early phases, one cannot rule out the role of binary mass transfer. A detailed stellar population synthesis to examine SN\,2021ukt as a binary progenitor system is beyond the scope of this paper. 

SNe\,2013df and 2020bio both show double-peaked light curves; the primary peak corresponds to the early shock-cooling emission of an extended outer envelope~\citep{soderberg2012panchromatic} and the secondary peak corresponds to radioactive decay of $^{56}\text{Ni}$. The progenitors of SNe\,2013df and 2020bio have a low $M_{\rm H,env}$:~\citet{10.1093/mnras/stu1837} found $M_{\rm H,env} \approx 0.2\,{\rm M}_{\odot}$ for SN\,2013df and~\citet{2023ApJ...954...35P} found $0.1\,{\rm M}_{\odot} \lesssim M_{\rm H,env} \lesssim 0.5\,{\rm M}_{\odot}$ for SN\,2020bio. The lack of broad \ion{H}{1} line profiles in the ejecta-dominated spectra of SN\,2021ukt effectively sets an upper limit on $M_{\rm H,env} \lesssim 0.5\,{\rm M}_{\odot}$. Furthermore, both SNe are found to have  $M_{\text{ZAMS}} \approx 12\,{\rm M}_{\odot}$~\citep{10.1093/mnras/stu1837,2023ApJ...954...35P}, consistent with our discussion in Section~\ref{sec:spec_nebu}.

Spectra of SN\,2021ukt exhibit a persistent emission of H$\alpha$ that is present from the CSM-dominated phase to the nebular phase. As shown in Figure~\ref{fig:h_vel}, we fit a Gaussian profile to the narrow Balmer emission lines during the first +5 days, measuring an initial FWHM $\approx 900$\,km\,$\text{s}^{-1}$. This measurement is comparable to previous measurements of SNe\,IIn \citep{1997ARA&A..35..309F,10.1093/mnras/stac1093}. 

The spectra during the ejecta-dominated phase then evolve to reveal conspicuous \ion{He}{1} P~Cygni profiles, similar to those of SNe\,Ib and IIb. The evolution of SN\,2021ukt from SN\,IIn to SN\,Ib is the first known transition of its kind (see \citealt{yesmin2025spectral}); if the spectrum at $t \approx 1$ month in the bottom panel of Figure~\ref{fig:spec_comp_ed} is actually closer to that of an SN\,IIb, then this would be the first example of an SN\,IIn to SN\,IIb transition. 
The ``shark-fin'' feature at 6400--6800\,\AA\ is likely a multicomponent emission line comprised of an intermediate-width H$\alpha$ component superimposed on the broad \ion{He}{1} $\lambda$6678 profile. SN\,IIb\,2013df followed an evolution from SN\,II to SN\,Ib-like, where its spectrum was initially observed with broad \ion{H}{1} P~Cygni profiles, which then evolved with weakening signatures of \ion{H}{1} that coincided with the prominence of broad \ion{He}{1} P~Cygni profiles~\citep{2014MNRAS.445.1647M}. The similarities in the evolution of SNe\,2013df and 2021ukt lead to our comparisons to SNe\,IIb.

After $\sim 6$ months, the H$\alpha$ line broadens to FWHM $\approx 2200$\,km\,$\text{s}^{-1}$. Meanwhile, H$\beta$ is no longer detected, likely owing to metal line blanketing~\citep{2020A&A...638A..80D}. Figure~\ref{fig:neb_spec} compares the nebular spectrum of SN\,2021ukt to nebular spectra of SNe\,IIb 1993J and 2008ax, SN\,Ib\,1990U, SN\,Ib/Ic 1990aj, and SN\,Ib--IIn 2014C. We observe relatively stronger [\ion{O}{1}] and [\ion{Ca}{2}] emission lines in the SNe\,IIb compared to SN\,2021ukt. These forbidden lines in typical nebular spectra of SNe\,II are comparably weaker than in SN\,2021ukt~\citep{2024MNRAS.528.3092L}. While the forbidden lines are relatively weaker in SN\,Ib\,1990U,~\cite{2015A&A...579A..95K} and~\cite{2019MNRAS.482.1545S} highlight how the [\ion{O}{1}] and [\ion{Ca}{2}] emission-line strengths of some SNe\,Ib are comparable to those of SNe\,IIb. We note that the observed blueshift in the line profiles presented in Figure~\ref{fig:ca_o_ratio} is likely due to inner layers of the expanding ejecta further heated by radioactive decay.

SN\,2014C exhibits an H$\alpha$ emission profile 113 days after its $V$-band maximum (FWHM $\approx 1200$\,km\,$\text{s}^{-1}$) that is most similar to that of SN\,2021ukt.~\cite{2015ApJ...815..120M} attributed this to delayed CSM interaction and noted that narrow Balmer emission (FWHM $\approx 250$\,km\,$\text{s}^{-1}$) from the host-galaxy \ion{H}{2} region also contributed to the overall line profile. CSM interaction could also be an explanation for the H$\alpha$ emission in the nebular spectrum of SN\,1990U (see Figure~\ref{fig:neb_spec}). Since SN\,2021ukt does not exhibit the constant narrow \ion{H}{1} emission line seen in SN\,2014C, contamination from the host galaxy is effectively ruled out. 

The early presence of conspicuous \ion{He}{1} and persistent narrow \ion{H}{1} emission results in the nebular-phase classification of SN\,2021ukt being consistent with both SNe\,Ib and IIn. In addition to the narrower width of H$\alpha$ compared to the [\ion{O}{1}] and [\ion{Ca}{2}] lines, we interpret the late-time evolution of the H$\alpha$ profile as evidence of ongoing interaction with extended, H-rich CSM. This CSM was possibly formed from H-rich material that was stripped from the progenitor star. Since some SNe\,Ib nebular spectra can show evidence of interaction (see SNe\,1990U and 2014C in Figure~\ref{fig:neb_spec}), the nebular-phase classification of SN\,2021ukt is consistent with an interacting SN\,Ib.  

The first panel of Figure~\ref{fig:aud_draw} illustrates the effect of shock interaction between optically thick, He-rich ejecta (red) interacting with a dense, nearby shell of CSM (blue, $\sim 10^{14}$\,cm). H$\alpha$ emission is likely due to the H-rich CSM photoionized by the initial shock wave, resulting in the Lorentzian-like line profile at day +3. The underlying continuum is represented by the approximately location of the expanding photosphere (yellow).

As represented in the middle panel of Figure~\ref{fig:aud_draw}, the emergence of the ``shark-fin'' feature at day +33 is interpreted as ongoing interaction between the He-rich ejecta and H-rich CSM. Figure 2 of \cite{yesmin2025spectral} presents this feature fitted with multicomponent Gaussian functions at +63 days to distinguish between the broad \ion{He}{1}\,$\lambda$6678 emission from the narrow \ion{H}{1} $\lambda$6563 emission; this is consistent with our interpretation of the ongoing shock interaction. Multicomponent emission lines were also identified in the spectra of SN\,IIn\,2020ywx~\citep{Baer-Way_2025}, which were found to be consistent with a system comprised of the expanding ejecta, shocked material, and unshocked, asymmetric CSM. While the multicomponent nature of the ``shark-fin'' feature could hint at CSM asymmetry, more detailed hydrodynamical modeling is needed to understand the mass-loss history.

The third panel of Figure~\ref{fig:aud_draw} illustrates the ejecta at day +180 continuing to interact with H-rich optically thin material ($\sim 10^{16}$\,cm), producing H$\alpha$ emission having FWHM consistent with  $v_{\text{CSM}} \approx 2200$\,km\,$\text{s}^{-1}$. The H$\alpha$ emission is likely due to the shock interaction between the optically thin ejecta and the extended CSM. 
 
\section{Summary and Conclusions}\label{sec:summ_conclu}
We have presented our analysis of the optical spectroscopy and photometry of SN\,2021ukt. The persistent H$\alpha$ emission indicates ongoing interaction with extended, H-rich material that could have been stripped from the progenitor prior to its explosion.

Our analysis is limited by several aspects, including poor constraints of the time of first light. Future observations of similar objects would greatly benefit from early-time UV observations, which would have helped constrain the progenitor environment that may influence the early-time light curve. Models capturing the complex physics associated with interaction with CSM or a binary companion may provide a better fit to the light curve.

We summarize the primary findings of SN\,2021ukt:

\begin{itemize}
    \item The evolution of SN\,2021ukt is the first known SN to spectroscopically transition from an SN\,IIn to an  interacting SN\,Ib (possibly SN\,IIb).

    \item Measurements of the [\ion{O}{1}] flux are consistent with a lower-mass progenitor, $M_{\text{ZAMS}} \approx 12\,{\rm M}_{\odot}$. Comparisons with SNe\,IIb also provide an upper limit of $M_{\rm H,env} \lesssim 0.5\,{\rm M}_{\odot}$.
        
    \item SN\,2021ukt is a member of a scarce population of SNe\,II having short-plateau light curves, 
    with an upper limit on the progenitor $M_{\text{Ni}}\lesssim0.04\,{\rm M}_{\odot}$ and $M_{\text{ZAMS}}\lesssim25\,{\rm M}_{\odot}$. 
    
    \item Given the fainter radioactive tail relative to other SNe and synthetic models, SN\,2021ukt may have a low $M_{\text{Ni}}$, or  low-density, high-velocity ejecta with inefficient trapping, or a combination of both conditions. 

    \item We interpret the lasting \ion{H}{1} emission as persistent interaction with an extended CSM, possibly formed from H-rich material that was stripped from the progenitor star prior to the explosion.
\end{itemize}

We emphasize that the progenitor of SN\,2021ukt is unique. Based on our analysis, the progenitor was likely an analog to an SN\,IIn with a lower ZAMS mass. The H-rich envelope that was stripped from the progenitor  remained sufficiently nearby to interact with the ejecta and produce relatively narrow emission lines. As the stripped material was swept up, the outer layers of the ejecta were revealed and evolved in close resemblance to those of SNe\,Ib with evidence of persistent CSM interaction. More modeling is needed to better understand the properties of the progenitor and its mass-loss history.

\section{Acknowledgments}

The authors dedicate this paper to the memory of William (Bill) Paxton, who created, developed, and maintained the highly valuable, open-source \texttt{MESA} package.
We are grateful for the UC Berkeley Mathematical and Physical Sciences Scholars Program and the Physics/Astronomy Undergraduate Research Scholars Program for funding this project. We acknowledge those who provided this project with their valuable help and insights: Wynn Jacobson-Galán, Karthik Yadavalli, Ryan Chornock, Raffaella Margutti, and Maryam Modjaz. In addition, we thank the students who made contributions to the Lick/Nickel photometric observations: Snehaa Kumar, Evelyn Liu, Emma McGinness, and Gabrielle Stewart.

A.V.F.’s research group at UC Berkeley received financial assistance from the Christopher R. Redlich Fund, as well as donations from Gary and Cynthia Bengier, Clark and Sharon Winslow, Alan Eustace and Kathy Kwan, William Draper, Timothy and Melissa Draper, Briggs and Kathleen Wood, Sanford Robertson (W.Z. is a Bengier-Winslow-Eustace Specialist in Astronomy, T.G.B. is a Draper-Wood-Robertson Specialist in Astronomy, 
Y.Y. was a Bengier-Winslow-Robertson Fellow in Astronomy), 
and numerous other donors. 
Y.Y.’s research is partially supported by the Tsinghua University Dushi Program.
K.A.B. is supported by an LSST-DA Catalyst Fellowship; this publication was thus made possible through the support of Grant 62192 from the John Templeton Foundation to LSST-DA.
The work of D.S. was carried out at the Jet Propulsion Laboratory, California Institute of Technology, under a contract with NASA  (\#80NM0018D0004).
This work makes use of observations from the Las Cumbres Observatory network. The LCO team is supported by NSF grants AST-2308113 and AST-1911151.
C.P.G. acknowledges financial support from the Secretary of Universities and Research (Government of Catalonia) and by the Horizon 2020 Research and Innovation Programme of the European Union under the Marie Sk\l{}odowska-Curie and the Beatriu de Pin\'os 2021 BP 00168 programme, from the Spanish Ministerio de Ciencia e Innovaci\'on (MCIN) and the Agencia Estatal de Investigaci\'on (AEI) 10.13039/501100011033 under the PID2023-151307NB-I00 SNNEXT project, from Centro Superior de Investigaciones Cient\'ificas (CSIC) under the PIE project 20215AT016 and the program Unidad de Excelencia Mar\'ia de Maeztu CEX2020-001058-M, and from the Departament de Recerca i Universitats de la Generalitat de Catalunya through the 2021-SGR-01270 grant.

Some of the data presented herein were obtained at the W. M. Keck Observatory, which is operated as a scientific partnership among the California Institute of Technology, the University of California, and the National Aeronautics and Space Administration (NASA); the observatory was made possible by the generous financial support of the W. M. Keck Foundation. 
A major upgrade of the Kast spectrograph on the Shane 3\,m telescope at Lick Observatory, led by Brad Holden, was made possible through gifts from the Heising-Simons Foundation, William and Marina Kast, and the University of California Observatories.
KAIT and its ongoing operation were made possible by donations from Sun Microsystems, Inc., the Hewlett-Packard Company, AutoScope Corporation, Lick Observatory, the U.S. National Science Foundation, the University of California, the Sylvia \& Jim Katzman Foundation, and the TABASGO Foundation.
Research at Lick Observatory is partially supported by a gift from Google.
We appreciate the expert assistance of the staff of the various observatories where data were obtained.

\bibliography{references}
\bibliographystyle{aasjournal}

\end{document}